\documentclass[onecolumn,traditabstract]{aa}
\usepackage{graphicx}
\usepackage{natbib}
\usepackage{epsfig}
\usepackage{lscape}
\usepackage{graphicx}
\usepackage{amsmath}
\usepackage[english]{babel}
\usepackage[titletoc,title]{appendix}
\usepackage{txfonts}
\usepackage{times}
\begin{document} 

\title{Observing the very low-surface brightness dwarfs in a deep field in the VIRGO cluster: constraints on  Dark Matter scenarios\thanks{Based on
observations made at the Large Binocular Telescope
(LBT) at Mt. Graham (Arizona, USA).}
}

\author{N. Menci, E. Giallongo, A. Grazian, D. Paris, A. Fontana, L. Pentericci
          }
\institute{INAF-Osservatorio Astronomico di Roma, via Frascati 33, 00078 Monteporzio, Italy
  }
   \date{}
   \authorrunning{}

\abstract
{We report the discovery of 11 very faint ($r\lesssim 23$),  low surface brightness ($\mu_r\lesssim 27$ mag/arcsec$^2$) dwarf galaxies in one deep field in the Virgo cluster, obtained by the prime focus cameras (LBC) at the Large Binocular Telescope (LBT). These extend our previous sample to reach a total number of 27 galaxies in a field of just of $\sim 0.17$ deg$^2$ located at a median distance of 390 kpc from the cluster center. Their association with the Virgo cluster is supported by their separate position in the central surface brightness - total magnitude plane with respect to the background  galaxies of similar total magnitude. For a significant fraction (26\%) of the sample the association to the cluster is confirmed by spectroscopic follow-up. We show that the mere abundance of satellite galaxies corresponding to our observed number in the target field provides extremely tight constraints on Dark Matter models with suppressed power spectrum compared to the Cold Dark Matter case, independently of the galaxy luminosity distribution. In particular, requiring the observed number of satellite galaxies not to exceed the predicted abundance of Dark Matter sub-halos yields a limit $m_X\geq 3$ keV at 1-$\sigma$ and $m_X\geq 2.3$ keV at 2-$\sigma$ confidence level for the mass of thermal Warm Dark Matter particles. Such a limit is competitive with other limits set by the abundance of ultra-faint satellite galaxies in the Milky Way, is completely independent of baryon physics involved in galaxy formation, and has the potentiality for appreciable improvements with next observations. We extend our analysis to Dark Matter models based on sterile neutrinos, showing that our observations set tight constraints on the combination of sterile neutrino mass $m_{\nu}$ and mixing parameter $sin^2(2\theta)$. We discuss the robustness of our results with respect to systematics.
}

   \titlerunning{Abundance of dwarf galaxies in a VIRGO field}
   \maketitle
%

\section{Introduction}

The abundance of dwarf satellite galaxies constitutes a key probe for Dark Matter (DM) scenarios. In fact, it strongly depends on the shape of the DM power spectrum at small galactic scales, which in turn is determined by the assumed mass of the DM candidates. In particular, the low velocity dispersion of Cold Dark Matter (CDM) particles results into large abundances of DM sub-halos around Milky Way-like galaxies which largely exceed (by a factor $\gtrsim 20$) the number of dwarf satellites observed around the Milky Way and M31 (Klypin et al. 1999, Diemand et al. 2008, Springel et al. 2008, Ferrero et al. 2012) 
so that a strong suppression of star formation in dwarf galaxies  (due to  feedback from both Supernovae and reionization at high redshift) must be invoked in order to suppress the number of luminous satellites compared to the abundance of DM sub-halos (see, e.g., Bullock,  Kravtsov, Weinberg  2000; Benson et al. 2002; Sommerville 2002;  Sawala et al. 2016). 
While the effectiveness of such processes in bringing the predicted abundance of dwarf galaxies in CDM in agreement with observations is still debeated (see, e.g., Schneider et al. 2016b), alternative DM models based on spectra with suppressed power on small scales have been proposed by several groups to solve the discrepancy. Among these, a  prominent class is constituted  by models based on lighter DM particles with masses in the keV range.  Compared to CDM particles, their larger thermal velocities suppress the growth of DM density fluctuations on small mass scales $M\lesssim 10^9$ $M_{\odot}$. Such DM candidates 
may be initially in thermal equilibrium (thermal warm dark matter models, WDM, see Bode, Ostriker, Turok 2001; De Vega, Sanchez 2010) or be produced from oscillations or decay of other particles, as in  sterile neutrino models (Colombi et al. 1995; see Adhikari 2017 for an extended review). In the former case, the suppression in the power spectrum with respect to CDM depends only on the exact value of DM particle mass $m_X$, since a thermalized species has no memory of the details of its production, while for sterile neutrinos the power spectrum depends on the production mechanism. E,.g., for sterile neutrinos produced by oscillations of active neutrinos, the power spectrum depends on both the 
mass $m_{\nu}$ and the mixing parameter $sin^2(2\theta)$. It is noticeable that oscillations make such DM candidates subject to direct observational test, since the decay of sterile neutrinos might result into the emission of X-ray photons at an energy $\approx 1/2\,m_{\nu}$. In fact, 
a tentative line signal at energy $\approx 3.5$ keV has already been reported in X- ray observations of stacked spectra of clusters of galaxies, as well as in several galaxy clusters (Bulbul et al. 2014; Boyarsky et al. 2014), in the X-ray emission of dwarf galaxies, of the Milky Way and of M31
(see, e.g., Sekiya et al. 2015; Jeltema, Profumo 2015, Riemer-S{\o}rensen 2014, Adhikari et al. 2017 for an extended discussion), and in the X-ray background (Cappelluti et al. 2017).

Since the key mass scales to distinguish CDM from WDM candidates are sub-galactic scales ($M\lesssim 10^9$ M$_{\odot}$), 
the abundance of low-mass dwarf galaxies has been extensively used to constrain both thermal WDM and sterile neutrino DM models.  In the case of thermal relic WDM particles, the one-to-one correspondence between the WDM particle mass and the suppression in the power spectrum at small scales has allowed to derive limits on $m_X$ by comparing the predictions from $N$-body WDM simulations or semi-analytic models with the abundance of observed ultra-faint satellites. On this basis, different authors have derived limits ranging from $m_X\geq 1.5$~keV (Lovell et al. 2014) to $m_X\geq 1.8$~keV (Horiuchi et al. 2013), $m_X\geq 2$~keV (Kennedy et al. 2013) and $m_X\geq 2.3$~keV (Polisensky, Ricotti 2010); relevant constraints have also been obtained for the parameter space of resonant production sterile neutrino models (Schneider 2016a), in that latter case also taking into account the actual shape of the distribution functions (and, thus, not using a WDM approximation to a non-thermal case). Note that, however, such methods are  sensitive to the assumed completeness corrections (see discussion in Abazajian et al. 2011, Schultz et al. 2014) and to the assumed values for the DM mass of the host halo and of the satellites. At higher redshifts, $z\approx 6$, a limit $m_X\gtrsim 1$~keV has been derived from the UV luminosity functions of faint galaxies ($M_{\rm UV}\approx -16$) in Schultz et al. (2014); a similar approach by Corasaniti et al. (2016) yield $m_X\gtrsim 1.5$~keV. Since these approaches are based on the comparison between the observed luminosity functions and the predicted mass function of DM halos in different WDM models, the delicate issue in these methods is their dependence on the physics of baryons, determining the mass-to-light ratio of faint galaxies. Although to a lesser extent, uncertainties in the baryonic physics  (related, e.g. to the average photoionization history of the IGM and to spatial fluctuations of the UV background) also affect (Garzilli et al. 2015; Viel et al. 2013; Ir\u{s}i\u{c} et al. 2017; see also Baur et al. 2017) the tighter constraints achieved so far  for WDM thermal relics, derived by comparing small scale structure in the Lyman-$\alpha$ forest of high-resolution ($z > 4$) quasar spectra with hydrodynamical $N$-body simulations (see Viel et al. 2013). The most recent results (Ir\u{s}i\u{c} et al. 2017) yield $m_X\gtrsim  5.3$~keV at 2-$\sigma$ confidence level from the combined analysis of two samples, although allowing for a non-smooth evolution of the temperature of the IGM can reduce the lower limit for the combined analysis to $m_X\geq 3.5$ keV (for a generalization of this method to sterile neutrinos models, see Schneider 2016a).

To bypass the uncertainties related to the physics of baryons affecting the constraints on DM models, Menci et al. (2016a,b) 
have exploited the downturn of the halo mass distribution  $\phi \left(M,z\right)$ in models with suppressed power spectra, which yields a maximum number density $\overline\phi$ of DM halos in the cumulative mass distribution that in turn depends on the adopted DM model. Since luminous galaxies cannot outnumber DM halos, an observed galaxy density ${\phi}_{obs}>\overline\phi$ would rule out the adopted DM model independently of the baryonic processes determining the luminous properties of galaxies. Such a method, first applied to lensed galaxies at $z=10$ in Pacucci et al. (2013)  and to galaxies at $z=7$ in the Hubble Deep Field in Lapi \& Danese (2015), has acquired an increased  potential with the first results of the Hubble Frontier Field (HFF) program. By exploiting the magnification power of gravitational lensing produced by foreground clusters, HFF  allows to reach unprecedented faint magnitudes $M_{\rm UV}=-12.5$ in the measurement of the luminosity function of galaxies at $z=6$ ( Livermore et al. 2017). The large number density ${\phi}_{obs}\geq 1.3\,{\rm Mpc}^{-3}$ (at 2-$\sigma$ confidence level) corresponding to the observed luminosity function allowed us to set a robust lower limit $m_X\geq 2.5$~keV (at 2-$\sigma$) to the mass of thermal relic WDM particles. This constitutes the tightest constraint derived so far on thermal WDM candidates independent of the baryon physics involved in galaxy formation. 

However,  systematics still affect the selection of highly magnified (typical amplifications $>10$ and as large as $\sim 50-100$ are involved), faint galaxies at high redshift, due to uncertainties related both to the lensing amplification and to the size distribution of faint galaxies. 
As for the lensing magnification, while the analysis in Livermore et al. (2017) shows that the uncertainties associated to the magnification of  individual galaxy is reduced appreciably when the aggregated information is considered, the proper procedure to be adopted in deriving the variance in the luminosity function due to the magnification is still matter of debate, see Bouwens et al. (2016a). 
As for the size distribution of high-$z$ galaxies, this strongly affects the completeness correction adopted in the estimates of the luminosity functions, since compact galaxies are more easily detected compared to the extended, low -surface brightness ones (Grazian et al. 2011). While in the simulations carried out by Livermore et al. (2017) a normal distribution of half-light radii $r_h$ has been assumed with a peak at 500 pc, assuming values $r_h\lesssim 100$~pc would lead to a significant suppression of the faint-end logarithmic slope $\alpha$ of luminosity function is found (up to 10\% in the case $r_h= 40$ pc, see Bouwens et al. 2016b). Maximizing the impact of both the above systematic effects would lead to looser constraints on the  DM models (see Menci et al. 2017 for a detailed discussion). 

Thus, it is important to complement the above high-redshifts baryon-independent constraints with independent low-redshifts limits derived from observables subjects to completely different systematics. This would also allow us to cross-check the bounds on the parameter space of DM models through observations of dwarf galaxies across a huge lapse of cosmic times. While previously mentioned works  mainly focused 
on Sloan Digital Sky Survey (SDSS) ultra-faint dwarf satellites of the Milky Way (Kennedy et al. 2013; Horiuchi et al. 2013; Lovell et al. 2014, Polisensky, Ricotti 2010), these still rely on the adopted corrections for the limited sky coverage of the SDSS, on the assumed mass of the Milky Way, and on the assumed lower limits for the mass of the dwarf galaxies. Such limits in turn  depend on the baryon physics entering either the luminosity-to-mass ratio of dwarf galaxies, or  the density profiles used to derive stellar velocity dispersion used to infer the dwarf galaxy masses. 
A complementary approach is to estimate the dwarf abundance in nearby clusters like Virgo or Coma. Searches in the Virgo cluster  allow us to probe a confined volume  in the local universe where the galaxy density is particularly high and where it is possible to select cluster members with high confidence. Cluster members  are easily selected  in the plane total magnitude - central surface brightness, 
since Virgo galaxies appear diffuse, with central surface brightness much fainter with respect to background  galaxies of similar total magnitude. Spectroscopic confirmation of photometric candidates has proved the reliability of this selection criterion (Rines \& Geller 2008). In addition, in the cluster environment a large dynamical range  $M_{sh}/M_h\sim 10^{-5}$  in the mass ratio  between the smaller subhalo and the main halo is easily obtainable without reaching the same very faint limits and low dwarf masses as in the MW halo. This large  range allows a more accurate statistical description of the subhalo distribution function together with a detailed investigation of the dynamical processes acting at the center of the cluster. 

In the present work we select low-surface brightness (LSB) dwarf galaxies in the Virgo cluster basing on deep LBT observations. We increase the Virgo dwarf sample we obtained in a previous paper (Giallongo et al. 2015) going deeper on a slightly larger area to evaluate a reliable lower limit to the total number density of faint dwarfs at a distance of about 390 kpc from the cluster center. Since our method is based on the 
comparison of the total abundance of satellites with the prediction of different DM models, we do not need to derive luminosity functions 
 or to make assumptions on the luminosity-to-mass ratio of the observed galaxies. Nevertheless, the mere increase in the observed abundance of galaxies in the target field compared to our earlier work,  when extrapolated to the whole cluster, will provide stringent limits on the parameter space of both the thermal WDM and the  sterile neutrino DM models. 
 
The paper is organized as follows. The data sample and the selection procedure are described in Section 2. The method we adopt to compare the 
 total abundance of galaxies in the observed field with the predictions of different DM models is presented in Section 3, where we also recall the 
  basic properties of the different DM scenarios we compare with. Our constraints on the DM models are presented in Section 4, while Section 5 is 
  devoted to discussion and conclusions. Throughout the paper we adopt round cosmological parameters  $\Omega_{\Lambda}=0.7$, $\Omega_{0}=0.3$, and Hubble constant $h=0.7$ in units of 100 km/s/Mpc. Magnitudes are in the AB photometric system.

\section{Selection of VIRGO galaxy candidates in the field}

In Giallongo et al. (2015, G15) we  exploited archival LBT images originally obtained by Lerchster et al. (2011) to select a sample of Virgo dwarf candidates over a $\approx 600$ arcmin$^2$ field located at $\sim 350$ kpc from the cluster center. We  detected 11 low surface brightness galaxies in the magnitude interval $-13\lesssim M_r \lesssim -9$. Such LSB galaxies are similar - although fainter in some cases - to those found in previous works by e.g. Sabatini et al. (2003), Lieder et al. (2012), Davis et al. (2015) and by Ferrarese et al. (2012,2016) on the basis of the NGVS survey at CFHT. The  latter  covers all the Virgo cluster for a total area of 100 deg$^2$ and the large statistics allows to derive a reliable faint-end slope of the Virgo luminosity function $\alpha \simeq -1.4$ (Ferrarese et al. 2016).

Since the goal of the present work is to derive an integrated lower limit to the number density of the dwarf Virgo galaxies we decided to go deeper in a relatively small area at an intermediate distance from the Virgo center as a complementary strategy to the surveys already completed in the overall Virgo cluster.  The selection criterion is based on the detection of the faintest dwarfs in terms of their central surface brightness.

As an intermediate step in this strategy we have obtained a deep Gunn-r image in a field partially overlapping with the previous one centered near the Virgo galaxy of the Markarian chain NGC4477 at an average distance of $\sim 390$ kpc. More specifically the lower and upper bounds in distance from the cluster center are at 310 kpc and 464 kpc, respectively. The distances have been estimated adopting the Virgo distance of 16.5 Mpc derived by Mei et al. (208) and a scale of about 80"/pc. Figure 1(left panel) shows a finding chart of the field together with the positions of the Virgo confirmed members and new candidates.
The new image has been obtained by the red channel of the Large Binocular Camera (Giallongo et al. 2008) at the prime focus of the 8m Large Binocular Telescope (LBT). The  final mosaic which includes the original image used in G15, has a total exposure time ranging from 9 to 11.5 ks over a field of view  of $\sim 0.17$ deg$^2$ centered near NGC4477. The image reaches a $2\sigma$ fluctuation in background surface brightness of the order of $\sim 28.6$ mag arcsec$^{-2}$ in the deepest area. The resulting average seeing corresponds to a  $FWHM\simeq 0.85$ arcsec.

The selection of Virgo galaxy candidates has been performed as in G15 exploiting the specific morphological properties of the Virgo galaxies in the central surface brightness - total magnitude plane, $\mu_0 -r$. This criterion has been adopted by several authors (Conselice et al. 2002, Rines \& Geller 2008), and its reliability has been extensively tested by spectroscopic follow up of  selected Virgo candidates.
More specifically, as in G15  we have used the SExtractor package (Bertin, Arnouts 1996) to select sources setting a brightness limit $\mu_r\simeq 27.5$ in a minimum detection circular area of diameter 2.5 arcsec. We have also used a circular aperture equivalent to 1 arcsec$^2$ as a first proxy of the central surface brightness of the diffuse sources. 

In Figure 1 (right panel)  we plot the galaxies found down to $r=23$ as a function of their central surface brightness. Only galaxies whose photometry is not  contaminated by other nearby sources are shown after a $\sigma$ clipping around the average linear regression.  The locus can be described by a linear relation of the type $\mu_0=0.94\times r+2.99$. At a first glance 27  galaxies are located well outside the locus of background galaxies  $\gtrsim 3\sigma$ away from the average locus of background galaxies. All 27 galaxies are clearly extended with low surface brightness and are selected as cluster members. Figure 1 also includes the 11 Virgo galaxies of G15. The candidates have been visually inspected to remove extended artifacts due to blending of background galaxies. As reported in G15 we emphasize that galaxies with characteristics similar to the selected Virgo dwarfs are absent in our database of even deeper LBC fields reaching limiting magnitudes $r\sim 26-27$ obtained in different sky regions far away from any nearby cluster, for a total area of 0.17 deg$^2$.

To better characterize the morphological and photometric properties of the Virgo candidates we have used the Galfit package (Peng et al. 2010) which performs radial profile modeling with S{\'e}rsic and other functions after convolution with the point-spread function profile as done in G15. This approach is  important in most cases where the LSB galaxy profile includes several overlapping compact sources which alter the estimate of LSB total magnitude and spatial profile. Thus small compact sources were simultaneously fit with the main LSB galaxy to remove their contribution.
The resulting best fit central brightness and total magnitudes are shown in Figure 1 for the Virgo dwaf candidates.  It is important to note that the faintest candidates in terms of central brightness are quite extended and not so faint in total magnitude. Thus shallower Virgo surveys could be affected by some incompleteness even at magnitudes significantly brighter than the survey limit. A detailed description  of the morphological and spectral characteristics of our sample is beyond the scope of the present work and will be published in a forthcoming paper.

Excluding the three brightest Virgo galaxies of the Markarian chain NGC 4477, NGC 4479 and NGC 4473  there are eight galaxies shown in Figure 1 which are present in the VCC catalog, four of which are confirmed Virgo members with spectroscopic redshift (VCC 1307, VCC 1351, VCC 1237, VCC 1198). Moreover, we have recently obtained spectroscopic follow up for 5 dwarfs of the G15 sample. Two of them have secure spectroscopic redshifts 
confirming they are part of the cluster, one has a probable redshift identification and the remaining two show too noisy spectra to derive a redshift (Giallongo et al., in preparation). Thus 26\% of the selected Virgo candidates from Figure 1 are spectroscopically confirmed Virgo members with no exception so far, supporting the reliability of the adopted morphological selection. Thus  for the analysis described in the next section we consider all the 27 galaxies as bona fide members of the Virgo cluster.

\begin{center}
\vspace{-0.1cm}
\hspace{-0.2cm}
\scalebox{0.35}[0.35]{\rotatebox{-0}{\includegraphics{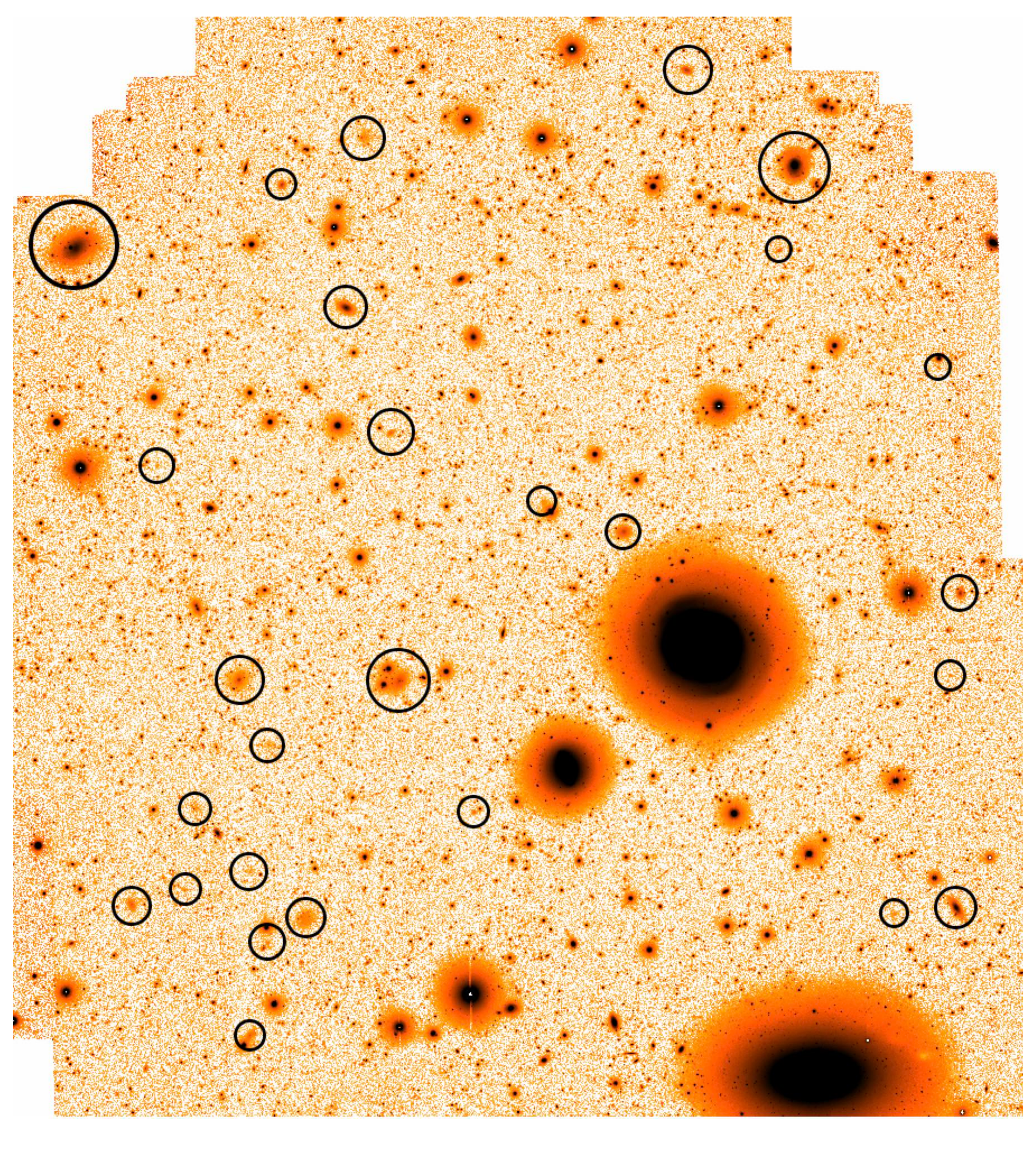}}}\hspace{1cm}
\scalebox{0.4}[0.4]{\rotatebox{-0}{\includegraphics{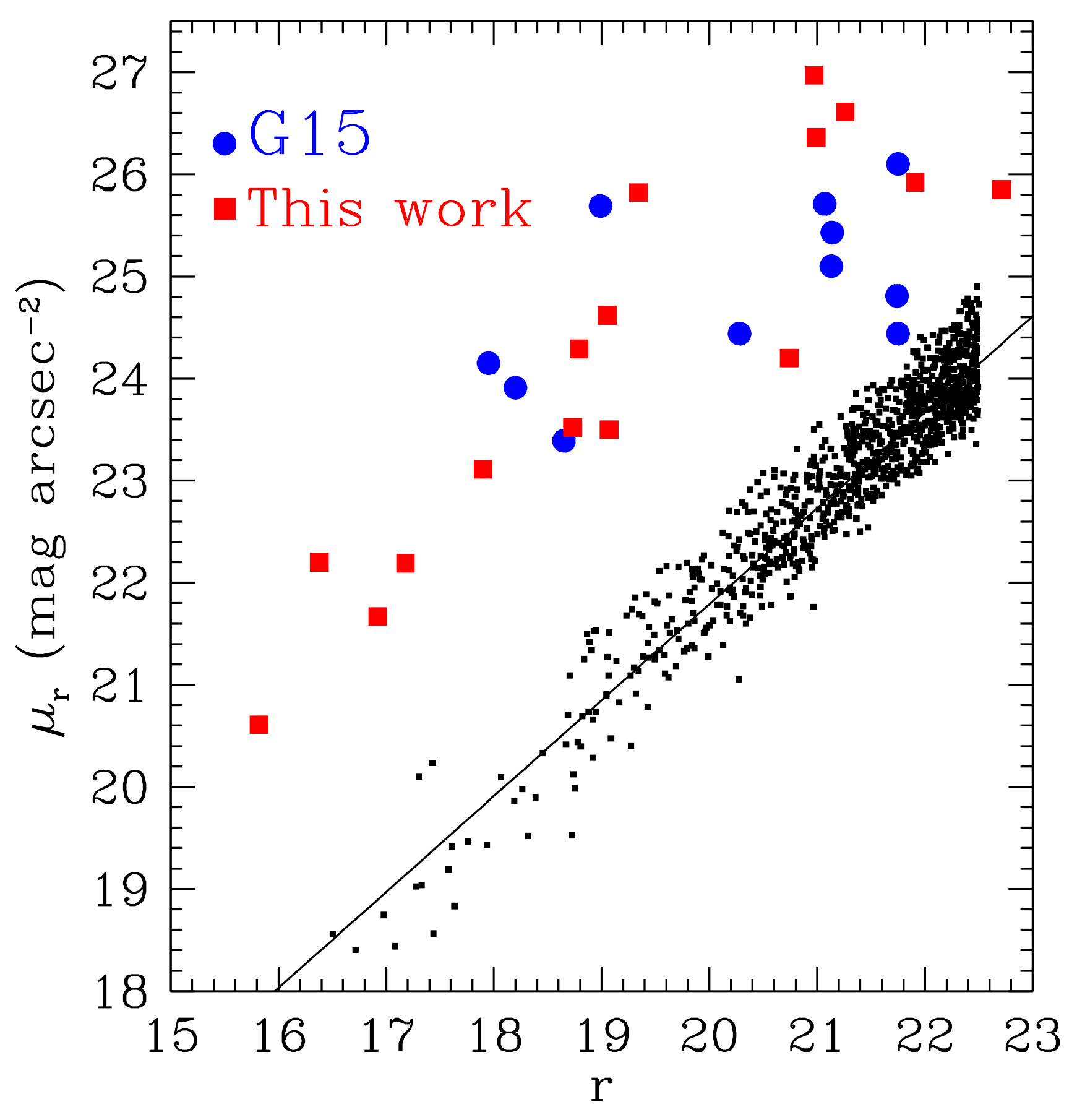}}}
\vspace{-0.2cm }
\end{center}
{\footnotesize Figure 1. Left Panel. Finding chart of our field around the bright galaxy NGC 4477.  The area is 0.17 deg$^2$. 
The image reaches a $2\sigma$ fluctuation in background surface brightness of the order of $\sim 28.6$ mag arcsec$^{-2}$.
\newline
Right Panel. Central surface brightness versus total Gunn-r magnitudes. The straight line represents the best fit of the correlation shown by the background galaxies (black points). Circles are LSB candidates selected in paper I (after zeropoint correction by -0.1 mag). Red large squares are new VIRGO candidates selected from the present catalog. }

\section{Abundance predictions from  Dark Matter scenarios and comparisons with the observed value}

The method is based on the drop of the differential sub-halo mass function $d\phi/dM_{sh}$ at small masses in DM models where the power spectrum is strongly suppressed  with respect to CDM at masses $M\lesssim 10^7-10^9\,M_{\odot}$. As a consequence, 
the corresponding cumulative sub-halo mass function $\int^{\infty}_{M_{sh}}\,dm ({d\phi/ dm})$ saturates to a maximum value $\overline{\phi}$ when the integral  is extended down to progressively smaller values of the satellite mass $M_{sh}$. The strength of the method is that it provides us with a maximum value for the number density of DM halos regardless of the underlying mass- luminosity relation, and - hence - completely independent of the baryon physics entering galaxy formation. 

The computation of the differential sub-halo mass function ${d\,\phi/ d\,logM_{sh}}$ in WDM models is based on the standard procedure 
described and tested against N-body simulations in Schneider et al. (2015, 2016, see also Giocoli, Pieri, Tormen 2008), derived in the framework of the Extended Press-Schechter approach. This can be expressed as: 
\begin{equation} 
{d\,\phi\over d\,logM_{sh}}={1\over C}\,{1\over 6\,\pi^2}\,{M_{hh}\over M_{sh}}\,{P(k)\over r^3\,\sqrt{\sigma_{sh}^2-\sigma_{hh}^2} }.
\end{equation}
The key quantity entering the computations is the  linear power spectrum 
$P(k)$ of DM perturbations, in terms of the wave-number $k=2\pi/r$, where $r$ is the spatial scale of DM density perturbations. 
Here we have used a sharp-$k$ form (a top-hat sphere in Fourier space) for the window function $W(kr)$ relating the variance to the power spectrum $\sigma^2(M)=\int dk\,k^2\,P(k)\,W(kr)/2\,\pi^2$ at the mass scale $M = 4\pi\,\overline{\rho} (q\,r)^3/3$ to the  filter scale $r$  (here $\overline{\rho}$ is the background density of the Universe).  The factor $q$ is calibrated through simulations; all studies in the literature yield values for $q$  in the range $q = 2.5-2.7$ (see, e.g., Angulo et al. 2013; Benson et al. 2013; Schneider et al. 2013) and we adopt the conservative value $q=2.5$. The normalisation constant takes the value $C=34$ when the host halo is defined as delimited by a density 200 times the critical density
inside a halo radius $R$. With such a definition, present observational uncertainties give for the Virgo cluster $M_{hh} =(4.5-5.5)\times 10^{14}\,M_{\odot}$ 
( McLaughlin 1999; Schindler, Binggeli, B\"ohringer 1999; Rines, Diaferio 2006; Urban et al. 2011; Durrell et al. 2014; Ferrarese et al. 2012, 2016).

Note that although eq. 1 has been tested against N-body simulations only in the case of CDM and thermal WDM models, the Extended Press \& Schechter formalism with sharp-$k$ filter on which it is based has been tested against simulations for a much wider set of DM power spectra. 
Recently Murgia et al. (2017) compared the mass functions computed in the Extended Press \& Schechter approach to high-resolution N-body simulations for a wide class of models with suppressed power spectrum compared to CDM, parametrized as to include the power spectra of sterile neutrino DM models (whether resonantly produced or from particle decay), finding a  good agreement between the theoretical mass function formalism outlined above and the N-body results. In fact, the above authors applied eq. 1 to compute the abundance of Milky Way satellites for the above class of DM  models.

For thermal WDM and sterile neutrino DM models, with power spectra suppressed at small scales compared to CDM, the  differential sub-halo mass functions 	(eq.1) are characterized by a maximum value  at masses close to the ‘half-mode’ mass (see Benson et al. 2013; Schneider et al. 2013; Angulo et al. 2013),  the mass scale at which the spectrum is suppressed by 1/2 compared to CDM. This is a strong inverse function of the sterile neutrino mass; for sterile neutrino models it  depends also on the assumed lepton asymmetry and, hence, on the resulting mixing angle $\theta$; typical power spectra in such models yield   half-mode masses ranging from $M_{hm}\approx 10^{10}\,M_{\odot}$ to $M_{hm}\approx 10^{8}\,M_{\odot}$.  Correspondingly, the cumulative mass functions saturate to a maximum value  $\overline{\phi}(z)\approx \phi(M_{hm},z)$, defining the maximum number density of DM halos associated to the considered power spectrum. 

For any given power spectrum $P(k)$ - corresponding to a chosen DM model - 
the cumulative sub-halo mass function $\phi(M_{sh})=\int^{\infty}_{M_{sh}}\,dm ({d\phi/ dm})$ determines the number of sub-halo with mass larger than $M_{sh}$ within the radius $R$. The corresponding maximum abundance of satellites within $R$ is $\overline{\phi}=\lim_{M_{sh}\to 0}\phi$. 
This has to be converted in an  expected number of galaxies in the observed field, which covers a 
region of the plane of the sky enclosed between $r_1=310$ kpc and $r_2$= 460 kpc from the cluster center, and extends for 120 kpc in the direction normal to the radius. Thus, the cumulative number of satellite galaxies (projected in the plane of the sky, normal to the line of sight) with mass larger than $M_{sh}$ expected in the observed field for a given DM model is given by 
\begin{equation} 
\phi_{field}(M_{sh},c,R)={\phi}(M_{sh})\,Q_1(c,R)\,Q_2
\end{equation}
where $Q_1$ is the fraction of galaxies in the circular region of the plane of the sky between $r_1$ and $r_2$, and $Q_2=120^2/[\pi\,(460^2-310^2)]=0.04$ is the fraction of the circular region covered by our field. The correction factor $Q_1$ depends on the assumed projected density distribution of galaxies in the cluster $\Sigma(r)$. For the latter we  assume the projection given by Lokas and Mamon (2001, eq. 41) of a NFW profile (Navarro, Frenk, White 1997) that is considered to produce an excellent fit for the surface distribution of galaxies in a large sample of clusters when  a concentration parameter $c=2.5$ is assumed (Budzynski et al. 2012). For the adopted profile, we compute the fraction of galaxies expected within a given projected radius $r$ as $F(<r)=\int_0^r \Sigma(r)\,r\,dr/\int_0^R \Sigma(r)\,r\,dr$. Then the factor $Q_1$ can be estimated as $Q_1=F(<r_2)-F(<r_1)$, and increases with increasing  $c$ and decreasing $R$. The uncertainties related to the computation of $Q_1$ are discussed in the next section.

Requiring the maximum number of DM satellites expected in a given DM model in the observed field to exceed (or to equal) the observed number of observed satellites $\phi_{obs}$ corresponds to the following condition:
\begin{equation}
\overline{\phi}_{field}(c,R)=\lim_{M_{sh}\to 0} \phi_{field}(M_{sh},c,R)\geq \phi_{obs}˜.
\end{equation}
On considering the statistical uncertainties on $\phi_{obs}$ and the uncertainties  on the density profile parameters $(c,R)$, Eq. (3) allows us  
to derive constraints on the DM model adopted to compute $\overline{\phi}_{field}$ within the proper confidence levels. Our results for WDM and for sterile neutrino DM are presented below. 

\section{Results: Constraining the Parameter Space of Dark Matter Models}

Here we presents the constraints that the observed number of Virgo satellites  in our field  (see Sect. 2) provides for 
WDM and sterile neutrino DM according to the method described above (Sect. 3). To this aim, we compute the power spectrum $P(k)$ for  different regions of the parameter space of WDM and sterile neutrino DM models, and we derive the corresponding satellite distribution after eq. (1). The corresponding cumulative mass distribution of satellites within our observed field is then derived for each chosen DM model from eq. (2). 

In eq. 2 we conservatively adopt for the DM mass of the Virgo cluster the value $M_{hh} =5.5\times 10^{14}\,M_{\odot}$ (the upper limit of the 
uncertainty range given in Sect. 3).  Note that adopting lower values for $M_{hh}$ would yield lower sub-halo abundances (see eq. 1) thus tightening our constraints.
As for the concentration parameter $c$ and  radius $R$ of the Virgo cluster, also entering Eq. (2), we 
 perform our analysis taking full account of the present observational uncertainties. As for the former, we shall consider the whole range $c=2-8$. In fact, while stacked analysis of a large sample of SDSS clusters shows that on average assuming $\langle c\rangle=2.6$ provides and excellent fit to the galaxy distribution inside many clusters (Budzynski et al. 2002), and the analysis  in McLaughlin (1999)  yields for Virgo $c=2.8\pm 0.7$, values up to $c=8$ have been reported (Simionescu et al. 2017) in the specific case of the Virgo cluster. For the 
 Virgo radius (defined as the region is enclosing an overdensity 200 with respect to the critical density), we adopt as a baseline the value  $R=1.6$ Mpc  following Ferrarese et al. (2012), but we shall consider the effects on our results of an uncertainty range $ R=1-3$ Mpc. 

For each assumed DM power spectrum $P(k)$, the above uncertainties in the parameters $c$ and $R$  provide an uncertainty range for the 
  projected number density of galaxies $\overline{\phi}_{field}$ expected in the observed field, that we compare (eq. 3) with 
 our observed value $\phi_{obs}$. To obtain robust constraints, we decrease our measured value ($\phi_{obs}=27$ galaxies, see Sect. 2) by 10\% to account for halo-to-halo variance (see, e.g., Mao, Williamson, Wechsler 2015). This yields a conservative central value estimate $\phi_{obs}=24.3$, with lower bounds $\phi_{obs}=19.37$, $\phi_{obs}=14.4$, and $\phi_{obs}=9.5$ at 1, 2, and 3-$\sigma$ confidence levels, respectively. Such lower bounds for the observed abundance of galaxies are compared (Eq. 3) with the expected value $\overline{\phi}_{field}$ computed for different regions of the parameter space of WDM and sterile neutrino DM models, to derive constraints with the corresponding confidence levels. 

\subsection{Thermal WDM scenario}
The only free parameter in such a scenario is the mass $m_X$ of the WDM particle. In fact, 
in this case the power spectrum can be parametrized as (Bode, Ostriker \& Turok 2001; see also Viel et al. 2005; Destri, de Vega, Sanchez 2013)
\begin{equation}
P_{WDM}(k)=P_{CDM}(k)\,\Big[1+(\alpha\,k)^{2\,\mu}\Big]^{-10/\mu}, 
\end{equation}
$${\rm ~with}~~~~~ 
\alpha=0.049 \,
\Bigg[{\Omega_X\over 0.25}\Bigg]^{0.11}\,
\Bigg[{m_X\over {\rm keV}}\Bigg]^{-1.11}\,
\Bigg[{h\over 0.7}\Bigg]^{1.22}\,h^{-1}\,{\rm Mpc},  \nonumber
$$
where $P_{CDM}$ is the CDM power spectrum (Bardeen et al. 1986), $\Omega_X=0.27$ is the WDM density parameter,  
$h$ is the Hubble constant in units of 100 km/s/Mpc, and $\mu=1.12$. 
The  ‘half-mode’ mass scale at which the WDM power spectrum is suppressed by 1/2 with respect to CDM is directly related to $m_X$  (see Schneider et al. 2012) by the relation $
M_{1/2}=(4\pi/ 3)\,\overline{\rho}\,[\pi\alpha(2^{\mu/5}-1)^{-{1/ 2\mu}}]^{3}$ 
where $\overline{\rho}$ is the background density of the Universe. For values of thermal relic masses $m_X=1.5-4$ keV considered here, one obtains $M_{1/2}= 10^8-10^9\,M_{\odot}$.  As shown by several  authors (Schneider et al. 2012, 2013; Angulo et al. 2013; Benson 2013), at this scale the DM mass function saturates and starts to turn off.

The number of sub-halos with mass $\phi(<M_{sh})$ expected in the observed field (eq. 1 and 2) is plotted in fig. 2 for power spectra corresponding to different WDM thermal relic mass $m_X$, and compared with the observed number (upper horizontal shaded region) corresponding to 24.3 galaxies (central value), 19 galaxies ($1-\sigma$ deviation), and 14.4 galaxies  ($2-\sigma$ deviation).
For each value of $m_x$, the thickness of the lines represents the uncertainty regions associated to profile parameters  discussed at the beginning of Sect. 2. For thermal relic masses $m_X<2.3$ keV the predicted number of satellites falls below the observed 
value at $2-\sigma$ confidence level, no matter how we extend to consider progressively smaller  satellite masses. This is clearly shown in the right panel of fig. 2, where the most conservative value for the maximum abundance of expected halos $\overline{\phi}_{field}$ is a function of the assumed WDM particle mass $m_X$ is compared with the observed values (at 1 and 2-$\sigma$ confidence level). 

\begin{center}
\vspace{0.1cm}
\hspace{-0.cm}
\scalebox{0.41}[0.41]{\rotatebox{-0}{\includegraphics{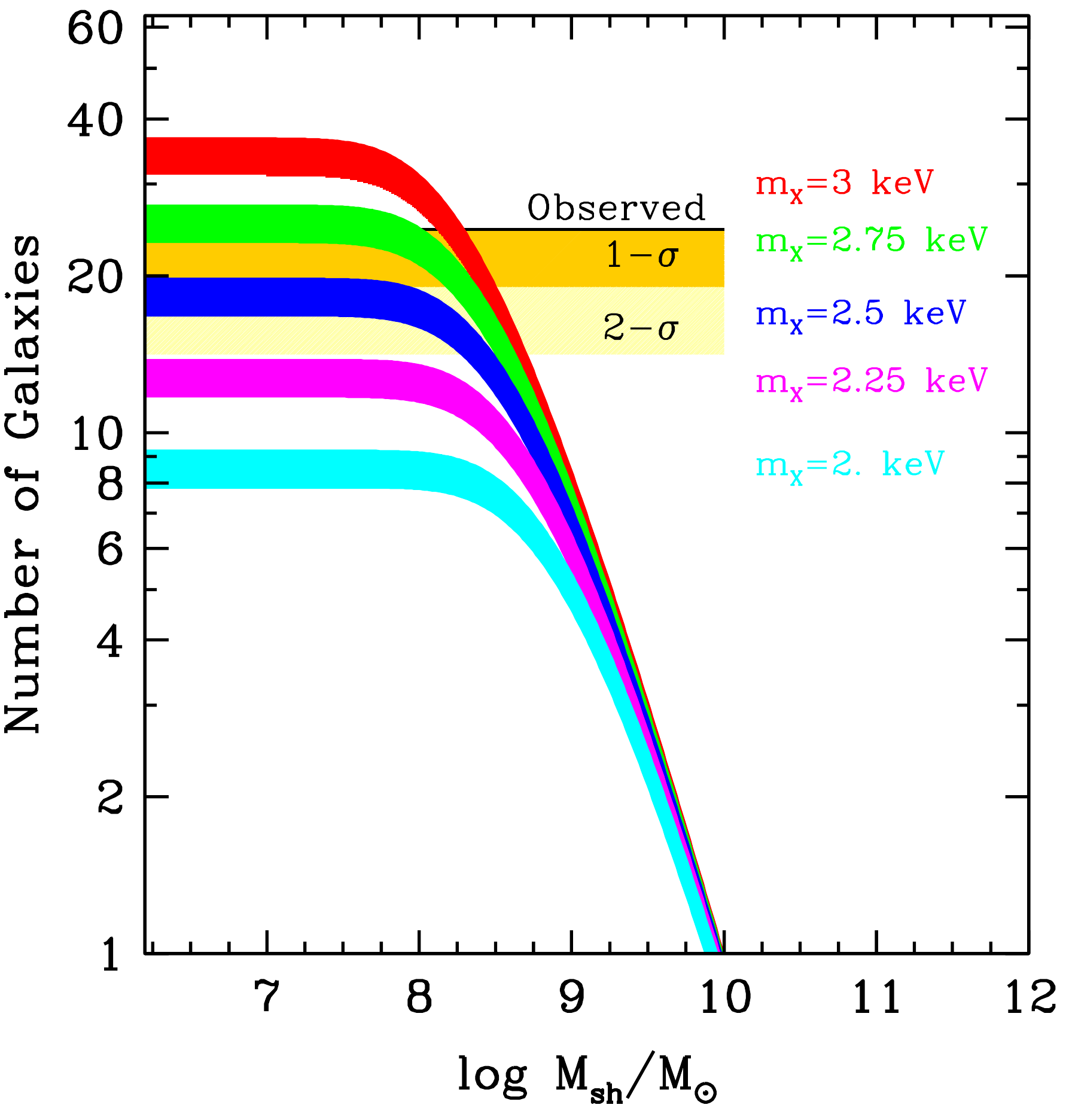}}}\hspace{1cm}
\scalebox{0.41}[0.41]{\rotatebox{-0}{\includegraphics{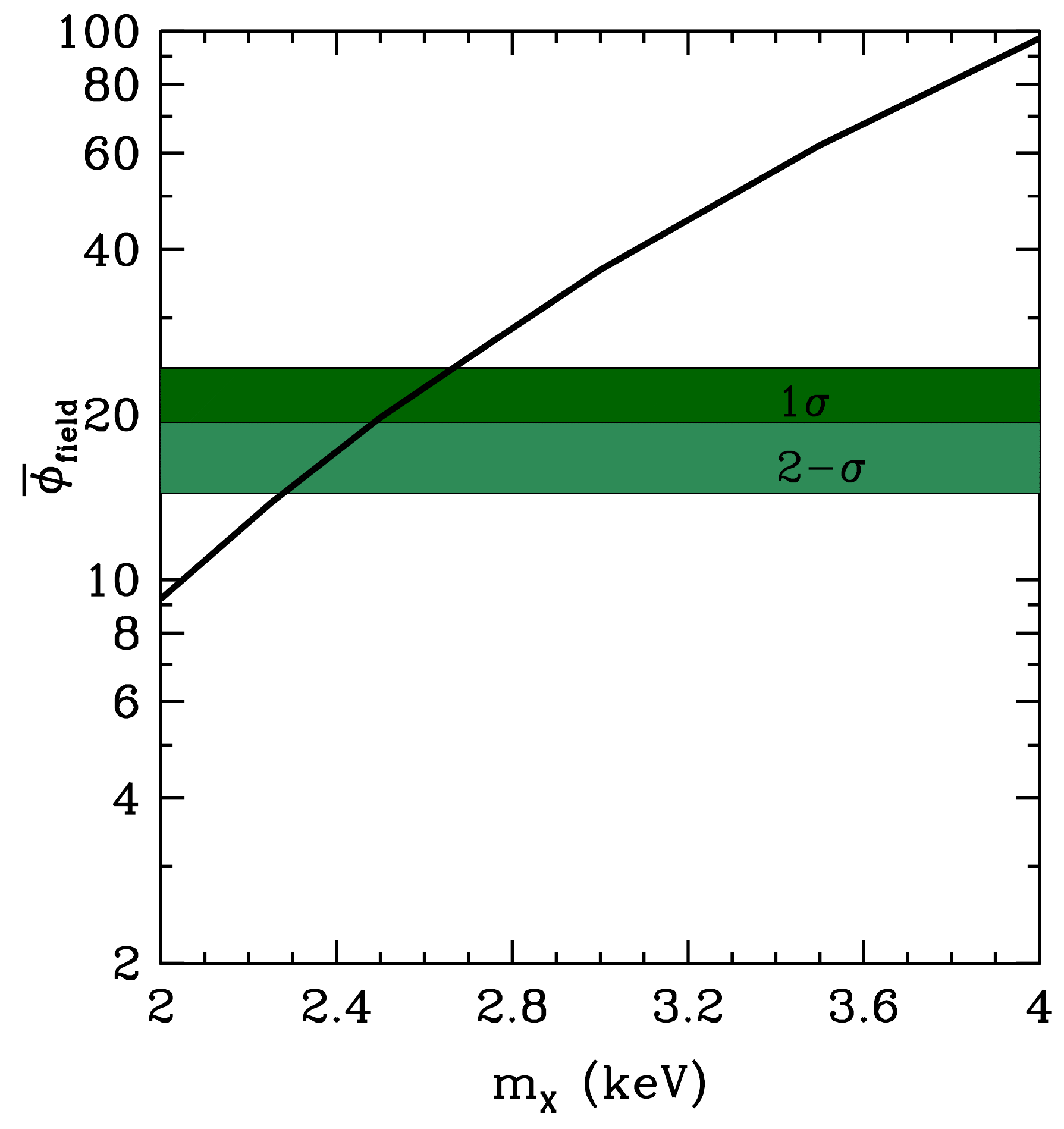}}}
\vspace{-0.cm }
\end{center}
{\footnotesize Figure 2. Left Panel. The cumulative number of sub-halos with mass $\phi(<M_{sh})$ expected in the observed field for WDM with different thermal 
relic mass shown in the labels. The  thickness of the lines represent the uncertainties in the theoretical predictions related to the uncertainties 
in the concentration parameter $c$ and in the radius $R$ assuming an uncertainty range  $1\leq c\leq 4$ and $R=1.5-2$ Mpc. The effect of extending the uncertainty ranges of $c$ and $R$ is shown in detail in fig. 3.
The predicted number is compared with the observed values $\phi_{obs}$ at 1-$\sigma$ and 2-$\sigma$ confidence levels, represented by the orizontal shaded regions. \newline
Right Panel. For different values of the thermal relic mass $m_X$, we show the maximum value of the predicted number of DM halos $\overline{\phi}_{field}$ in the observed field, and compare it with the observed number  within 1-$\sigma$ and 2-$\sigma$   
confidence levels.}
\vspace{0.3cm}

Our constraints are slightly tighter than with previous bounds $m_X\gtrsim 1.5-2$ keV found though the abundance of Milky Way ultra-faint galaxies 
(e.g., Kennedy et al. 2013; Horiuchi et al. 2013; Lovell et al. 2014, Polisensky, Ricotti 2010), but obtained with observations affected by different systematics. While our constraints are independent on baryon physics, on the assumed luminosity-to-mass ratio of the satellite galaxies, and on their dynamical mass, they are affected by uncertainties in the parameters $c$ and $R$ entering the density profile of the galaxy distribution in the Virgo cluster. However, our constraints on $m_X$ are characterized by a slow dependence on such parameters, as shown in fig. 3. Indeed, increasing $c$ (the parameter most subject to observational uncertainty) up to the value $c=8$ will affect our constraints by less than 5 \%, while assuming values $R>1.6$ for the virial radius would lead to even tighter bounds on $m_X$. 

\begin{center}
\vspace{-0.1cm}
\hspace{-0.8cm}
\scalebox{0.52}[0.52]{\rotatebox{-0}{\includegraphics{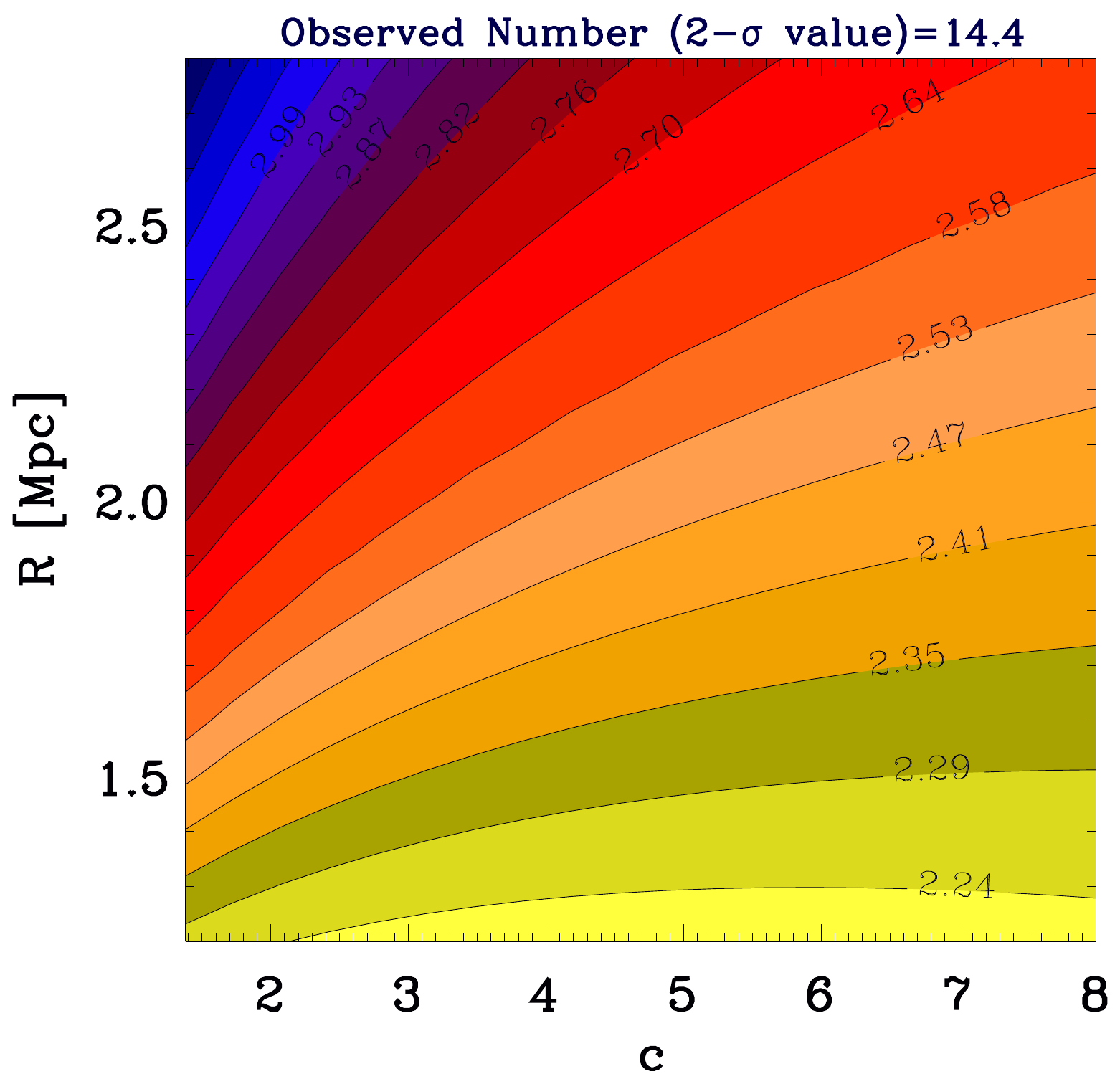}}}
\end{center}
\vspace{-0.3cm}
{\footnotesize 
Figure 3. We show how our 2-$\sigma$ limits on the thermal relic mass $m_X$ (coloured contours) depend on the asssumed values for the concentration parameter $c$ (horizontal axis) and for the cluster  radius $R$ (vertical axis).}
\vspace{0.4cm}

In addition, the constraints presented here can be  appreciably tightened with aimed observations in the next future. In fig. 4 we show how the constraints on the thermal WDM particle mass can be improved by increasing the observed value (2-$\sigma$ confidence level) for different values of the concentration parameter $c$. This can be achieved either going deeper (thus increasing the central value $\phi_{obs}$), or enlarging the field area 
(thus reducing the variance at fixed $\phi_{obs}$). E.g., doubling the number of observed galaxies at fixed  field area to reach a number of galaxies  $\phi_{obs}=50$ (so that at $\phi_{obs}=35$ at 2-$\sigma$ confidence level) would enable to constrain WDM models with $m_X\approx$ 3 keV. For a luminosity function steeper than $L^{-1.5}$ this would require a deepening of about 1.5 mag with respect to the present observations. 

\begin{center}
\vspace{-0.1cm}
\hspace{-0.8cm}
\scalebox{0.5}[0.5]{\rotatebox{-0}{\includegraphics{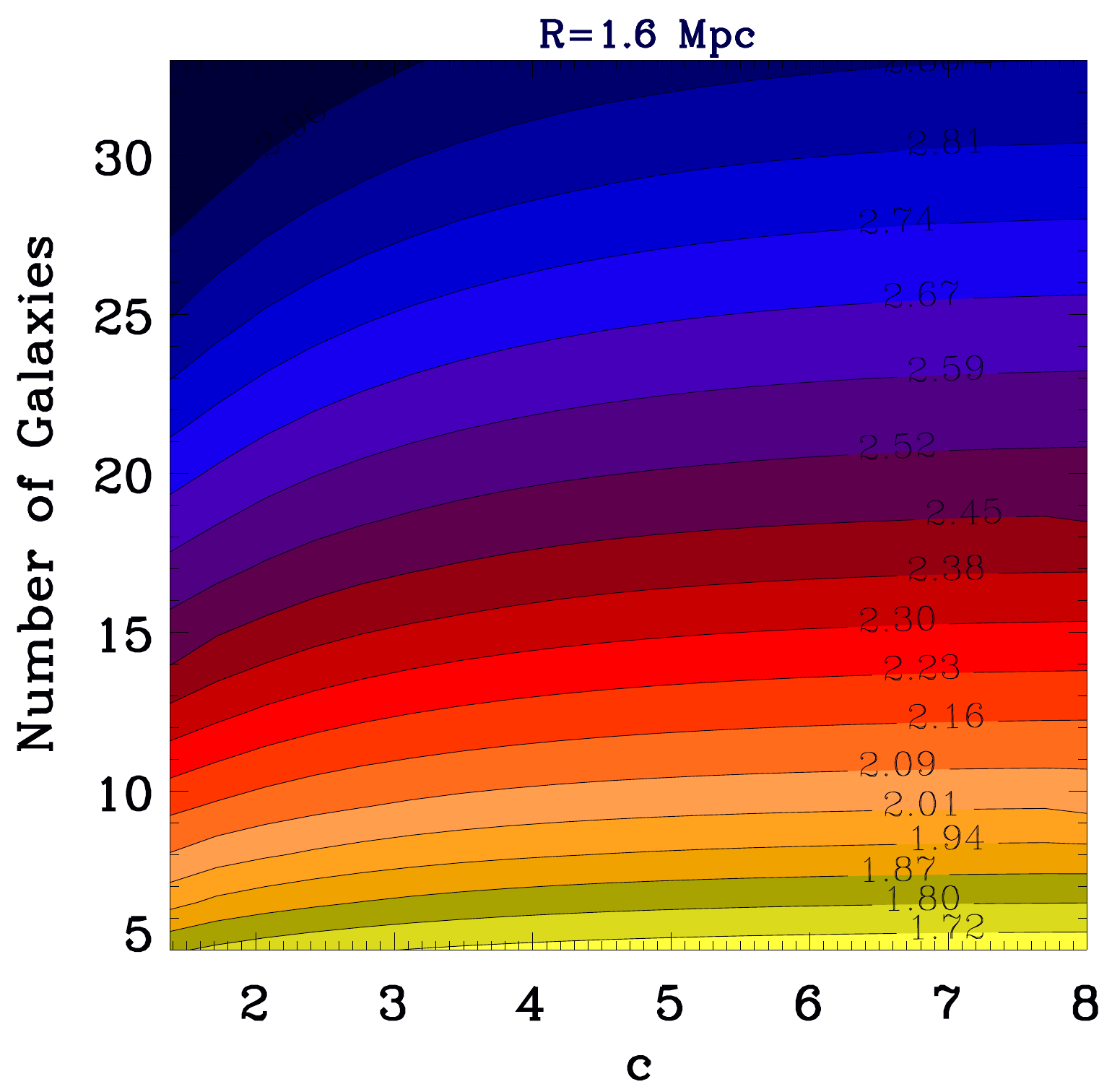}}}
\end{center}
{\footnotesize 
Figure 4. We show how our 2-$\sigma$ limits on the thermal relic mass $m_X$ (colored contours) change if the observed value $\phi_{obs}$ (at 
 2-$\sigma$ confidence level, vertical axis) is increased over the value measured in the present paper ($\phi_{obs}$(2-$\sigma$)=14.4), assuming 
 for the concentration parameter the different values shown in the x-axis. Here $R=1.6$ Mpc is assumed as a reference value. 
 }

\subsection{Sterile Neutrino DM}

In this scenario the power spectrum entering Eq. 1 depends not only on the mass $m_{\nu}$ but also on the production mechanism. Here we shall consider only the production via occasional oscillations from active neutrinos in the early Universe (Dodelson, Widrow 1993; for alternative scenarios based on the decay of scalar particles see Merle \& Totzauer 2015 and references therein).  In this case, for any given sterile neutrino mass, the mixing angle  $\theta$  depends on the assumed lepton asymmetry $L$ in the early Universe. In the limit $L\rightarrow 0$ (usually referred to as Dodelson-Widrow limit) the corresponding mixing angles are too large to comply with existing limits from  from X-ray observations (see, e.g., Canetti et al. 2013; Boyarsky et al.al. 2014;  Adhikari et al. 2017). However, in the presence of a significant lepton asymmetry $L$ in the early Universe, sterile neutrino production can be enhanced by a resonance (Shi, Fuller 1998), thus allowing for significantly smaller mixing angles $\theta$ which are consistent with present X-ray bounds. 
Since, for any given sterile neutrino mass, the mixing angle is related to the adopted lepton asymmetry $L$, in this work we describe the parameter space of sterile neutrino models in terms of combinations of sterile neutrino masses $m_{\nu}$ and mixing amplitudes $sin^2(2\theta)$. Each one of such combinations results in a density perturbation power spectrum $P(k)$ which is differently suppressed on small scales compared to CDM. 
In this work the sterile neutrino momentum distributions for resonant production have been computed with the public code {\it sterile-dm} of Venumadhav et al. (2015). To obtain the power spectra, the publicly available Boltzmann solver {\it CLASS} (Blas, Lesgourgues, Tram2011;  Lesgourgues, Tram 2011) has been used.

We explore the whole range of free parameters using a grid of values for both $m_{\nu}$ and $\sin^2(2\theta)$. After computing the corresponding power spectra, the condition $\overline{\phi}_{field}\geq {\phi}_{obs}$ (Eq. 3) leads to the exclusion region in the plane $m_{\nu}-\sin^2(2\theta)$ shown in Fig. 5. This is computed for the reference case $c=3$ and $R=1.6$ Mpc. However, the sensitivity of the curves to changes in such parameters 
is the same of the thermal WDM case (fig. 3) so that our constraints change by less than 5\% when such parameters are allowed to vary in the range $c=2.4-8$ and $R=1-3$. 

While non-resonant DW limit $L=0$ (the upper green curve in the left panel of Fig. 5) is excluded by X-ray observations, for lower values of $\sin^2(2\theta)$ allows to comply with the X-ray observations and - at the same time - yields a number of sub-halos large enough to be consistent with the observed values $\phi_{\rm obs}$, thus resulting into the allowed region of Fig. 5. When the lepton number is further increases, i.e., for even smaller values of $\sin^2(2\theta)$ warmer power spectra are obtained, thus yielding the lower exclusion region. 

\begin{center}
\vspace{-0.1cm}
\hspace{-0.8cm}
\scalebox{0.45}[0.45]{\rotatebox{-0}{\includegraphics{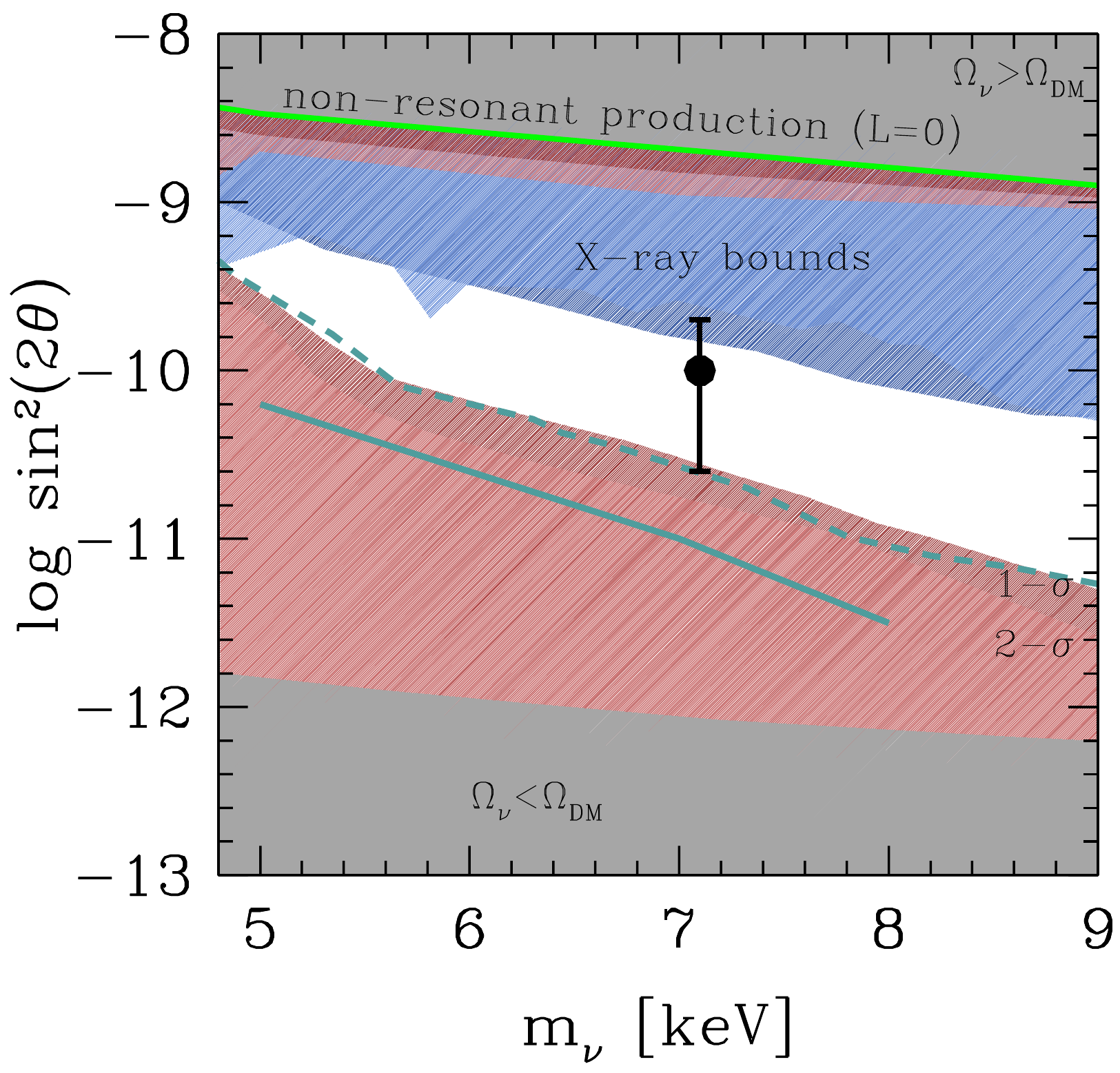}}}
\scalebox{0.43}[0.43]{\rotatebox{-0}{\includegraphics{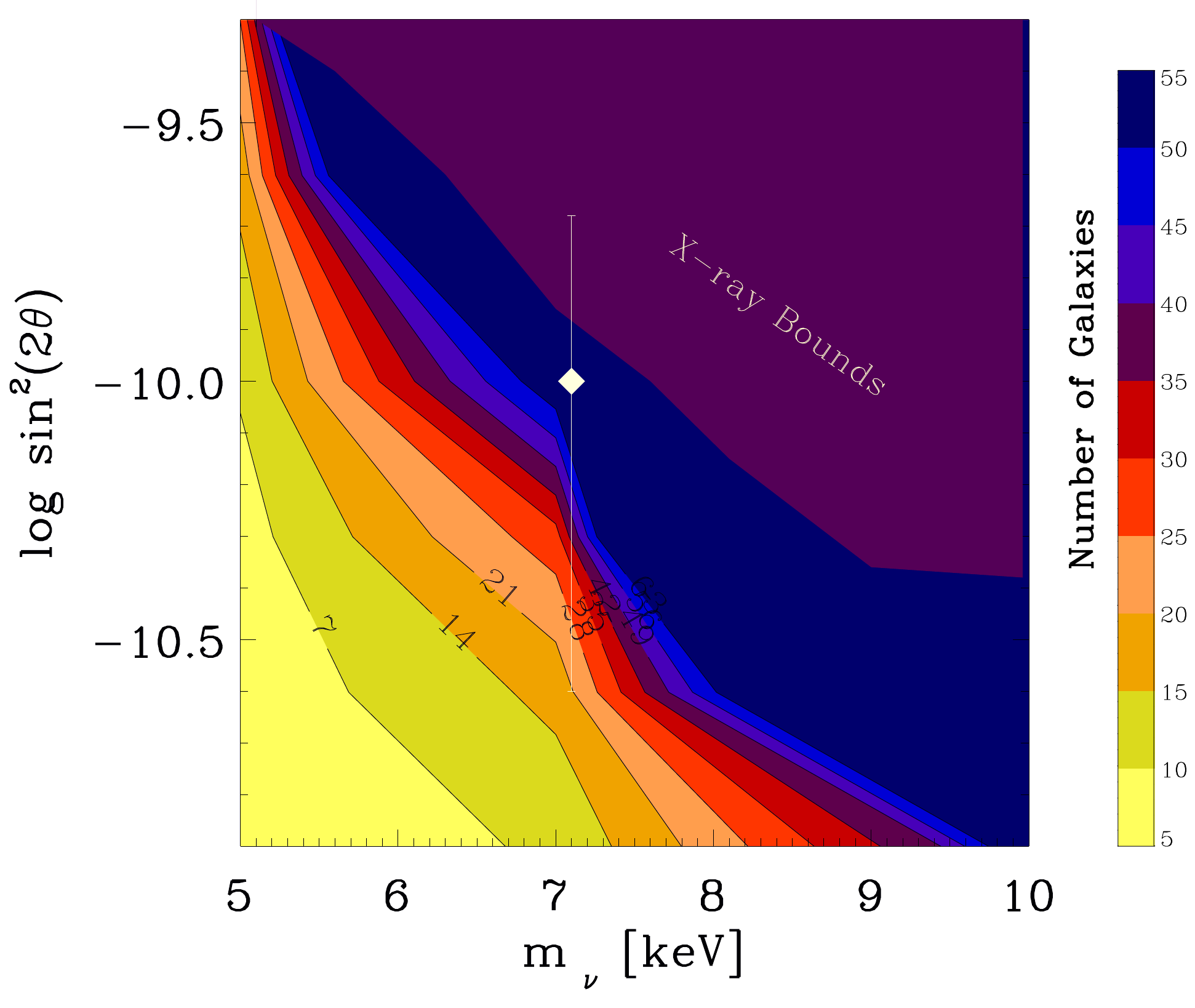}}}
\end{center}
{\footnotesize  Fig. 5. Left Panel: The constraints on the  sterile neutrino parameter space from Eq. (3). The allowed region is the white area. 
Taking the abundance $\phi_{obs}$ corresponding to our observed number of galaxies in the target field within 1-$\sigma$ and  2-$\sigma$ confidence levels ($\phi_{obs}=19.3$ and $\phi_{obs}=14.4$, respectively) yields the exclusion regions shown as red shaded areas.  Our constraints are compared with upper bounds from X-ray observations  of the Milky Way and dwarf galaxies (as reported in Riemer-S{\o}rensen 2014). Grey areas are excluded by current limits on the abundance of DM (Ade et al. 2015). The green line  corresponds to the non-resonant DW case with vanishing lepton asymmetry $L=0$. We also show  the constraints obtained from the abundance of satellites in the Milky Way by Schneider (2016a, dashed line) and by Lovell et al. (2016, solid line).  
The tentative line signal at 3.5~keV corresponding to a sterile neutrino mass $m_{\nu}$=7.1 keV (Bulbul et al. 2014;  Boyarsky et al. 2014) is shown by the point with errorbars.  \newline
Right Panel: The colored contours correspond to the different values of $\overline{\phi}_{field}$ corresponding to a given combination of $\sin^2(2\theta)-m_{\nu}$. \newline 
In both panels we assumed $c=3$ and $R=1.6$ Mpc as reference values. The sensitivity of the curves to changes in such parameters 
is the same of the thermal WDM case, shown in fig. 3.}
\vspace{0.3cm}

We note that our observations lead to a mass limit $m_{\nu}\geq 5$ keV for the sterile neutrino independently of the mixing angle. For larger masses, 
 the measured number of galaxies in our target field yields strong lower limits for the allowed range of $sin^2(2\theta)$. In particular, for neutrino masses  $m_{\nu}=7$ keV our observations leave allowed a region $log sin^2(2\theta)\geq -10.2$; interestingly, the tentative $3.5$ keV line signal reported in several works based on X-ray observations (Cappelluti et al. 2017; Bulbul et al. 2014; Boyarsky et al. 2014j; Sekiya et al. 2015; Jeltema, Profumo 2015;  Riemer-S{\o}rensen 2014; Adhikari et al. 2016) 
 falls within the range allowed by our results and by existing upper X-ray bounds. We also show the bounds on the $sin^2(2\theta)-m_{\nu}$ plane 
 obtained from the abundance of ultra-faint Milky Way satellites by Schneider et al. (2016a) and Lovell et al. (2016). Our  results match almost perfectly  those in Schneider et al. (2016a). This is particularly interesting, since the constraints from Milky Way satellites have been obtained through methods subject to  different systematics compared to our analysis, which include assumed lower limits for the satellite mass (obtained either through the luminosity-to-mass ratio or the measurement of stellar velocity dispersions). On the other hand, Milky Way constraints  based on satellite abundances computed through a semi-analytic computation  (Lovell et al. 2016) found a somewhat looser limits. 

Finally, in the right panel of fig. 5 we show the full distribution of the expected number of galaxies in the target field $\overline{\phi}_{field}$ corresponding to each combination of $\sin^2(2\theta)-m_{\nu}$. This can be used as a guideline for future observations, since measuring 
 abundances of satellite galaxies $\phi_{obs}=\overline{\phi}_{field}$ would correspond to probing the portion of the $\sin^2(2\theta)-m_{\nu}$ plane 
 marked by the corresponding contour. E.g., observing a number $\phi_{obs}=50$ galaxies in our field would lead to probe the parameter space deep 
 within the present errorbar of the tentative 3.5 kev line. 
 
\section{Conclusions}

We have extended our previous sample of low-surface brightness galaxies in the Virgo cluster measuring a total number $\phi_{obs}=27$ galaxies 
in a narrow field of $\sim 0.17$ deg$^2$ located at a median distance of 390 kpc from the cluster center. We have shown that the corresponding total galaxy population of satellite galaxies in the Virgo cluster provides strong constraints on DM models with suppressed power spectra compared to CDM. 
In particular, requiring the  total number of satellite galaxies in the Virgo cluster not to exceed the maximum number of DM sub-halos provides 
limits $m_X\geq 2.3$ keV (at 2-$\sigma$ confidence level) for thermal WDM models, and provides robust lower limits on the mixing 
parameter  $\sin^2(2\theta)$ of sterile neutrino DM models for each  value of the sterile neutrino mass $m_{\nu}$. In particular, for DM models with sterile neutrino masses $m_{\nu}=7$ keV (those possibly associated to the tentative 3.5 keV line) our observations yield a lower limit 
$log sin^2(2\theta)\geq -10.2$.The above limits can be translated into bounds on the lepton asymmetry $L_6
\equiv 10^6(n_{\nu_e}-n_{\overline{\nu}_e})/s$, 
defined in terms of the difference in electron neutrino and anti electron neutrino abundance divided by
the entropy density. For a given mass $m_{\nu}$, this is inversely related to the mixing parameter  (see, e.g., fig. 1 in Lovell et al. 2015;  Boyarsky et al. 2009 and references therein). E.g., for sterile neutrinos with mass $m_{\nu}=7$ keV our constraints on the mixing parameter correspond to $L_6 \leq $ 10.

We have shown that such bounds are independent on baryon physics entering galaxy formation. Indeed,  our results do not require any measurement 
of the  properties of the dwarf galaxies (like their DM or stellar mass, their luminosity or colors, their luminosity distributions) but only rely on the 
observed abundance of satellite dwarfs in the target field. Indeed, while we performed a detailed investigation of the robustness of our results with respect to the statistical uncertainties and to the systematics affecting the measured dwarf satellite abundance, 
we postpone a detailed investigation of the stellar and luminous properties of such galaxies in a next paper. 

On the other hand, our bounds on the WDM and sterile neutrino DM models are competitive with existing limits 
from the abundance of Milky Way satellites, and with those arising from the abundance of faint galaxies at $z=6-7$ detected in the HFF program. 
This is particularly important since the consistency with bounds derived from observations involving completely different scales of host halo mass (in the case of Milky Way satellites) or of cosmic times (in the case of high-redshift galaxies) shows a convergence of existing limits on DM from cosmic structures to the same values ($m_X\gtrsim 2.5$ keV for thermal WDM particles). This is even more noticeable when one considers the completely different systematics that affects the various methods. In fact, limits derived from the abundance of Milky Way satellites are affected by assumptions concerning their spatial distributions, by measurements used to derive their DM mass, and by the assumed mass of the Milky Way; systematics affecting the bounds derived from the abundance of high-redshift are related to the magnification of such objects and to  the assumed size-luminosity relation 
(Bouwens et al. 2016); and the abundance of satellites of the Virgo cluster presented in this paper depends on assumed the galaxy spatial distribution 
as described in Sect. 3 and 4. Nevertheless, when such aspects are properly addressed, all methods converge to provide similar constraints on the parameter space of DM models. In the case of sterile neutrino DM, it is also noticeable that all methods yield bounds that are consistent with the decay origin of the tentative 3.5 keV line in the X-ray spectra of different objects. 

We stress that the bounds on the DM properties based on the abundance of dwarf satellites in the Virgo cluster presented in this paper are  easier to be significantly improved compared to other constraints based on the abundance of dwarf galaxies. 
 In fact, at present uncertainties in the magnification and sizes limit the improvement of bounds from faint, high-redshift galaxies, while the 
  the detection of very diffuse structures like ultra-faint dwarfs in the Milky Way will be very challenging due to the contamination by foreground stars, and will involve very deep observations over significantly large portions of the sky, a task that will require LSST (which will observe half of the sky down to $r\sim 27$ approximately by the year 2028, see Ivezic et al. 2016). Instead,  
 deeper observations of the galaxies in the Virgo cluster have the potentiality to improve rapidly to reach unprecedented baryon-independent constraints on the  mass of thermal WDM particles and on sterile neutrino DM models.  

In fact, such a potentiality is witnessed by  the improvement of the new observations in the present paper with respect to the 
 the results reported in Giallongo et al. (2015) and originates from the combined increase in the detection threshold of both magnitude and surface brightness.  Indeed, inspection of Figure 1 shows that new LSB galaxies have been found  both at $r>22$ and at $\mu_r>26$. 
 This shows that tighter limits can be achieved with reasonable efforts in next LBT observations of the Virgo cluster by combining deeper observations with larger areas. E.g., covering 4 LBC FoVs  to reach a $2\sigma$ fluctuation in background surface brightness of the order of $\sim 30$ mag arcsec$^{-2}$ would require approximately 120 hours with LBT in the next 2-3 years. If the Virgo luminosity function extends 1.5 magnitudes fainter with the same faint-end slope $\simeq -1.4$, this will allow us to find about $\sim 50$ dwarfs (with a central surface brightness $\lesssim 28.5$ mag arcsec$^{-2}$) within a 0.15 deg$^2$ area,  with half the present variance. This  would  yield a lower bound $\phi_{obs}\geq 50-7=43$  at 2 $\sigma$ confidence level. This (if the presence of fainter galaxies will be confirmed by future observations)
would allow us to exclude at 95.4\% c.l. (i.e. 2$\sigma$) the value
$m_X\approx 3.3$ keV (see fig. 2, right panel) for the thermal relic case (comparable to that set by the Lyman-$\alpha$ forest, but independent of baryon physics), and would restrict the allowed range of values for  mixing angle of sterile neutrino models to an unprecedented narrow region (see fig. 5, right panel), so as to probe the region consistent with the tentative 3.5 keV line. 
 
It is interesting to note that this program can be executed only with
wide field instruments at 8-10m class telescopes. For comparison,
JWST, thanks to the low background from space, will reach deeper
magnitudes, but will cover significantly smaller areas of the sky, so
it will not be able to carry out such a survey (although it will improve the
 limits from ultra faint, strongly lensed high-redshift  galaxies, see Menci et al. 2017).

\begin{acknowledgements}

Observations have been obtained using the Large Binocular Telescope at Mt. Graham, AZ. The LBT is an international collaboration among institutions in the United States, Italy, and Germany. LBT Corporation partners are The University of Arizona on behalf of the Arizona university system; Istituto Nazionale di Astrofisica, Italy; LBT Beteiligungsgesellschaft, Germany, representing the Max-Planck Society, the Astrophysical Institute Potsdam, and Heidelberg University; The Ohio State University; and The Research Corporation, on behalf of The University of Notre Dame, University of Minnesota, and University of Virginia. 

\end{acknowledgements}

\end{document}